\documentstyle[12pt,epsf,amsmath,amsfonts]{article}

\begin{document}
   

\begin{center}

 {\Large \bf
\vskip 7cm
\mbox{Azimuthal Angular Distributions in EDDE as}
\mbox{spin-parity analyser and glueball filter for LHC.}
}
 
\vskip 1cm

\vskip 3cm

\mbox{Petrov~V.A., Ryutin~R.A., Sobol~A.E.}

\mbox{{\small Institute for High Energy Physics}}

\mbox{{\small{\it 142 281} Protvino, Russia}}

\vspace*{0.3cm}
and
\vspace*{0.3cm}

\mbox{J.-P. Guillaud}

\mbox{{\small LAPP, Annecy, France}}

 \vskip 1.75cm
{\bf
\mbox{Abstract}}
  \vskip 0.3cm

\newlength{\qqq}
\settowidth{\qqq}{In the framework of the operator product  expansion, the quark mass dependence of}
\hfill
\noindent
\begin{minipage}{\qqq}

Exclusive Double Diffractive Events (EDDE) are analysed as the source of information about the central system. Experimental possibilities for exotic particles searches are considered. From the reggeized tensor current picture some azimuthal angle dependences were obtained to fit the data from WA102 experiment and to make predictions for LHC collider.

\end{minipage}
\end{center}


\begin{center}
\vskip 0.5cm
{\bf
\mbox{Keywords}}
\vskip 0.3cm

\settowidth{\qqq}{In the framework of the operator product  expansion, the quark mass dependence of}
\hfill
\noindent
\begin{minipage}{\qqq}
Exclusive Double Diffractive Events -- Pomeron -- Regge-Eikonal model -- glueball -- Higgs
\end{minipage}

\end{center}

\setcounter{page}{1}
\newpage


\section{Introduction}

Since the early time the process of exclusive production of central systems of particles with quasi-diffractively scattered initial particles was considered as an important source of information about high-energy dynamics of strong interactions both in theory and experiment.

If one takes only one particle produced, this is the first "genuinely" inelastic process which not only retains a lot of features of elastic scattering but also shows clearly how the initial energy is being transformed into the secondary particles. General properties of such amplitudes were considered in Ref.~\cite{Logunov}.

Theoretical consideration of these processes on the basis of Regge theory goes back to papers~\cite{2}. Some new interest was related to possibly good signals of centrally produced Higgs bosons and heavy quarkonia~\cite{3}.

As Pomerons are the driving force of the processes in question it is naturally to expect that glueball production will be favorable, if one believes that Pomerons are mostly gluonic objects~\cite{4}. Central glueball production was suggested as possible origin of the total cross section rise in Ref.~\cite{5}. One of the early proposal for experimental investigations of centrally produced glueballs in EDDE was made in~\cite{Prokoshkin}.

As to the most recent experimental studies one has to mention the series of results from the experiment WA102~\cite{WA102}-\cite{WA102b}.

It was proposed in Ref.~\cite{6} that the process of single-particle diffractive production can serve as a filter to separate $q\bar{q}$-states from glueballs due to special dependence on azimuthal angle.

In this paper we study the process of single particle production in double diffractive events in the framework of Regge picture (based on Lorentz tensor reggeized exchanges) with account of absorbtion effects both in initial and final state. With parameters fixed from the fit of the WA102 data we give predictions for production of various $J^{PC}$ states at LHC. The main conclusion from WA102 that $q{\bar q}$-states and glueballs have distinct $\phi$-dependence (with account of possible mixing) remains true, in our approach, for LHC energy, though the functional form of $\phi$-distributions changes due to significant absorbtion effects.


\section{EDDE kinematics and cross-sections.}

Here we consider the process $p+p\to p+X+p$, where X is a particle or a system of particles,
spin and parity of which are fixed. In the kinematics which corresponds to the double Regge limit 
(see Fig.1) the light-cone representation $(+,-;\bot)$ for momenta of colliding and scattered particles is the following: 
\begin{eqnarray}
\label{momenta}
p_1&=&\left( \sqrt{\frac{s}{2}},\frac{m^2}{\sqrt{2s}}, {\bf 0}\right)\\
p_2&=&\left( \frac{m^2}{\sqrt{2s}},\sqrt{\frac{s}{2}}, {\bf 0}\right)\nonumber\\
p_1^{\prime}&=&\left( (1-\xi_1)\sqrt{\frac{s}{2}},\frac{{\bf\Delta}_1^2+m^2}{(1-\xi_1)\sqrt{2s}}, -{\bf\Delta}_1\right)\nonumber\\
p_2^{\prime}&=&\left( \frac{{\bf\Delta}_2^2+m^2}{(1-\xi_2)\sqrt{2s}},(1-\xi_2)\sqrt{\frac{s}{2}}, -{\bf\Delta}_2\right).\nonumber
\end{eqnarray}
$\xi_{1,2}$ are fractions of protons' momenta carried by reggeons. The boldface type is used 
for two-dimensional transverse vectors. From the above notations we can
obtain the relations:
\begin{eqnarray}
\label{notations}
t_{1,2}&=&\Delta_{1,2}^2 \simeq -\frac{{\bf\Delta}_{1,2}^2(1+\xi_{1,2})+\xi_{1,2}^2m^2}{1-\xi_{1,2}}\simeq -{\bf\Delta}_{1,2}^2\;{,}\;\;\xi_{1,2}\to 0\\
\cos\phi_0&=&\frac{{\bf\Delta}_1{\bf\Delta}_2}{|{\bf\Delta}_1||{\bf\Delta}_2|}\nonumber\\
M_{\bot}^2&=&\xi_1\xi_2s\simeq M_X^2+|t_1|+|t_2|+2\sqrt{t_1t_2}\cos\phi_0\;, 0\le\phi_0\le\pi\nonumber\\
(p_1&+&\Delta_2)^2=s_1\simeq \xi_2 s\nonumber\\
(p_2&+&\Delta_1)^2=s_2\simeq \xi_1 s\;\nonumber\\
s_1 s_2&=&s M_{\bot}^2\nonumber
\end{eqnarray}
Physical region of double diffractive events is defined by the following  
kinematical cuts:
\begin{equation}
\label{tlimits}
0.01\; GeV^2\le |t_{1,2}|\le 1\; GeV^2\;{,} 
\end{equation}
\begin{equation}
\label{xlimits}
\xi_{min}\simeq\frac{M_X^2}{s \xi_{max}}\le \xi_{1,2}\le \xi_{max}=0.1\;.
\end{equation} 
The discussion on the choice of cuts~(\ref{tlimits})-(\ref{xlimits}) for diffractive events 
and references to other authors were given in~\cite{Collins},\cite{Cudell}. We 
can write the relations in terms of $y_{1,2}$ and $y_X$ (rapidities of hadrons and the system X correspondingly). 
For instance:
\begin{eqnarray}
\label{raplimits}
&\xi_{1,2}&\!\!\simeq\frac{M_X}{\sqrt{s}}e^{\pm y_X}\; ,\\
&|y_X|&\!\!\le y_0=\ln\left(\frac{\sqrt{s}\xi_{max}}{M_X}\right)\; ,\nonumber\\
&|y_{1,2}|&\!\!=\frac{1}{2}\ln\frac{(1-\xi_{1,2})^2s}{m^2-t_{1,2}}\ge 9\nonumber
\end{eqnarray}

The cross-section of this process can be written as
\begin{equation}
\label{xsec}
\frac{d\sigma}{dt_1dt_2d\phi_0 dy_X}\simeq\frac{\pi |T^{Unit.}_{pp\to pXp}|^2}{8s^2(2\pi)^5}\;,
\end{equation}
where $T^{Unit.}_{pp\to pXp}$ is the amplitude of the process, which can be
calculated from the "bare" amplitude of Fig.1 by the unitarization
procedure depicted in Fig.2, where 

\begin{eqnarray}
\label{ucorr}
T_X&=&T_{pp\to pXp}\;,\\
V(s\;,\;\mbox{\bf q}_T)&=&4s(2\pi)^2\delta^2(\mbox{\bf q}_T)
+4s\int d^2\mbox{\bf b}e^{i\mbox{\bf q}_T \mbox{\bf b}}
\left[e^{i\delta_{pp\to pp}}-1\right]\;,\nonumber\\
T^{Unit.}_X(p_1\;,\; p_2\;,\;\Delta_1\;,\;\Delta_2)
&=&\frac{1}{16ss^{\prime}}\int
\frac{d^2\mbox{\bf q}_T}{(2\pi)^2}\frac{d^2\mbox{\bf q}^{\prime}_T}{(2\pi)^2}
V(s\;,\;\mbox{\bf q}_T)\;\cdot\nonumber\\
&\cdot&T_X( p_1-q_T, p_2+q_T,\Delta_{1T},\Delta_{2T})
\cdot V(s^{\prime}\;,\;\mbox{\bf q}^{\prime}_T)\;,\nonumber\\
\Delta_{1T}&=&\Delta_{1}-q_T-q^{\prime}_T\;,\nonumber\\
\Delta_{2T}&=&\Delta_{2}+q_T+q^{\prime}_T\;,\nonumber
\end{eqnarray}
and $\delta_{pp\to pp}$ can be found in Ref.~\cite{Petrov3P}. Left and right parts $V$ represent "soft" rescattering
effects for initial and final states, i.e. multi-Pomeron exchanges. These "outer" unitarity corrections
can reduce the integrated cross-section 
by the factor, which is determined by kinematical cuts and the nature of the produced system X. 

\section{Calculation of the "bare" amplitude.}

In order to calculate peculiar momentum transfers and azimuthal angle dependence we use the amplitudes with meson exchanges of arbitrary spins with subsequent reggeization.

 Basic elements of such approach are vertex functions
 \begin{equation}
 \label{vertexI1}
 T^{\mu_1\dots\mu_J}(p,\Delta)=<p-\Delta| I^{\mu_1\dots\mu_J}|p>
 \end{equation}
 and
 \begin{eqnarray}
 \label{vertexI2}
 &&F^{\mu_1\dots\mu_{J_1},\;\nu_1\dots\nu_{J_2}}_{\alpha_1\dots\alpha_J}(\Delta_1,\Delta_2,p_X)=\int d^4 x\;d^4 y\; {\bf e}^{ -i \Delta_1 x-i \Delta_2 y}\cdot\\
 &&\cdot <0| T^* I^{\mu_1\dots\mu_{J_1}}\left(x\right)
 I^{\nu_1\dots\nu_{J_1}}\left(y\right)
 I_{\alpha_1\dots\alpha_J}\left(0\right)|0>\;,\nonumber
 \end{eqnarray}
where  $I^{\mu_1\dots\mu_{J}}$ is the current operator related to the hadronic spin-$J$ field operator,
\begin{equation}
\label{KGJ}
\left( \square + m_J^2\right) \Phi^{\mu_1\dots\mu_J}(x)=I^{\mu_1\dots\mu_J}(x)\;.
\end{equation}

The amplitude $T_{pp\to pXp}$ (Fig.1) is composed of vertices $T^{\mu_1\cdots\mu_{J_1}}$, $T^{\nu_1\cdots\nu_{J_2}}$, $F^{\mu_1\dots\mu_{J_1},\;\nu_1\dots\nu_{J_2}}_{\alpha_1\dots\alpha_J}$ and propagators $d(J,t)/(m^2(J)-t)$ which have the poles at
\begin{equation}
\label{polesJ}
m^2(J)-t=0,\; \mbox{i.e.}\; J=\alpha(t)\;,
\end{equation}
after an appropriate analytic continuation of the signatured amplitudes in $J$. We assume that these poles, where $\alpha$ are Pomeron trajectories, give the dominant contribution at high energies after having taken the corresponding residues. Regge-cuts are generated by unitarization.

 For vertex
functions $T_{1,2}$ we can obtain the following tensor decomposition:
\begin{eqnarray}
\label{ppJvertex}
&&T^{\mu_1\dots\mu_J}(p,\Delta)=T_0(\Delta^2) \sum_{n=0}^{\left[\frac{J}{2}\right]} Y_J^n {\cal T}_{00J,\;n}^{\mu_1\dots\mu_J}\;,\\
&&Y_J^n=\frac{2^n (2(J-n))\mbox{!}J\mbox{!}D_p^{2n}}{(J-n)\mbox{!}(2J)\mbox{!}}\\
&&D_p^{\;\mu}=2p^{\mu}-\Delta^{\mu},\; D_p^2=4 m^2-\Delta^2,
\end{eqnarray}
that satisfies Rarita-Schwinger conditions:
\begin{eqnarray}
\label{RaritaShwinger}
&&T^{\mu_1\dots\mu_i\dots\mu_j\dots\mu_J}=T^{\mu_1\dots\mu_j\dots\mu_i\dots\mu_J}\label{symmetric}\\
&&\Delta_{\mu_i}T^{\mu_1\dots\mu_i\dots\mu_J}=0\label{transverse}\\
&&g_{\mu_i\mu_j}T^{\mu_1\dots\mu_i\dots\mu_j\dots\mu_J}=0\label{traceless}
\end{eqnarray}
Tensor structures ${\cal T}_{00J,\;n}^{\mu_1\dots\mu_J}$ satisfy only two 
conditions~(\ref{symmetric}),(\ref{transverse}) and consist of the elements $D_p^{\mu}$ and $G^{\mu\nu}$:
\begin{equation}
\label{Gmunu}
G^{\mu\nu}=-g^{\mu\nu}+\frac{\Delta^{\mu}\Delta^{\nu}}{\Delta^2}\;.
\end{equation}
\begin{equation}
\label{trstensor}
{\cal T}_{00J,\;n}^{\mu_1\dots\mu_J}=D_p^{\; \left(\mu_1\right.}\cdot\dots\cdot D_p^{\; \mu_{J-2n}}
G^{\mu_{J-2n+1}\mu_{J-2n+2}}\cdot\dots\cdot G^{\left.\mu_{J-1}\mu_{J}\right)} 
\end{equation}

Let us assume the fusion of two particles with spins $J_1$ and $J_2$ into a particle with spin $J$. The
general structure 
$F^{\mu_1\dots\mu_{J_1},\;\nu_1\dots\nu_{J_2}}_{\alpha_1\dots\alpha_J}(\Delta_1,\Delta_2,p_X)$ (see Fig.1) 
should satisfy conditions~(\ref{symmetric})-(\ref{traceless}) on each group of indices. Since
the contraction of the vertex with structures $T^{\mu_1\dots\mu_{J_1}}(p_1,\Delta_1)$,
$T^{\nu_1\dots\nu_{J_2}}(p_2,\Delta_2)$ and polarization tensor $e^{\alpha_1\dots\alpha_J}(p_X)$ of the X particle 
leads to vanishing of some terms in $F$, the remainder can
be constructed from 
\begin{equation}
\label{constrtensor}
p_X^{\mu_i},\; p_X^{\nu_j},\; \Delta_1^{\alpha_k}\; 
(\mbox{or}\; \Delta_2^{\alpha_k}),\; g^{\mu_i\nu_j},\; 
g^{\mu_i\alpha_k},\; g^{\nu_j\alpha_k} 
\end{equation}
for tensors and additional terms
\begin{eqnarray}
\label{constrpseudo}
&&\Lambda_X^{\mu_i\nu_j\alpha_k}=p_X^{\rho}\epsilon^{\rho\mu_i\nu_j\alpha_k},\\
&&\Lambda_n^{\mu_i\nu_j\alpha_k}=\Delta_n^{\rho}\epsilon^{\rho\mu_i\nu_j\alpha_k},\\
&&Q_n^{\lambda\kappa}=\Delta_n^{\rho}p_X^{\sigma}\epsilon^{\rho\sigma\lambda\kappa}\to
\Delta_1^{\rho}\Delta_2^{\sigma}\epsilon^{\rho\sigma\lambda\kappa},\\ 
&&n=1\;\mbox{or}\; 2,\; (\lambda\kappa)=(\mu_i\nu_j),\;(\mu_i\alpha_k),\; (\nu_j\alpha_k)\nonumber\\
&&i\le J_1,\; j\le J_2,\; k\le J\nonumber
\end{eqnarray}
for pseudo-tensors~\cite{Vertex3J}. Let $J_1\le J_2$ and consider several cases. 
\begin{itemize}
\item $J^{P}=0^{+}$
\begin{eqnarray}
\label{J0p}
&&F^{\mu_1\dots\mu_{J_1},\;\nu_1\dots\nu_{J_2}}(\Delta_1,\Delta_2,p_X)=\sum_{k=0}^{J_1}
f_k\cdot\\
&&\cdot \left( p_X^{\mu_1}\cdot\dots\cdot p_X^{\mu_{J_1-k}}\cdot 
g^{\mu_{J_1-k+1}\nu_{J_2-k+1}}\cdot\dots\cdot g^{\mu_{J_1}\nu_{J_2}}\cdot 
p_X^{\nu_{J_2-k}}\cdot\dots\cdot p_X^{\nu_1}\right)\;.\nonumber
\end{eqnarray}
After tensor contraction we obtain the expansion of the type
\begin{equation}
\label{legendre}
\sum_{n=0}^{\left[\frac{J_i}{2}\right]}\frac{(-1)^nC_{J_i}^n C_{J_i}^{2n}}{C_{2J_i}^{2n}}
\left(
\sqrt{\frac{\xi_i^2(m^2-t_i/4)}{-t_i}}
\frac{M_X^2-t_1-t_2}{M_{\bot}^2}
\right)^{2n}\;.
\end{equation}
In the kinematical region~(\ref{tlimits}),(\ref{xlimits})
$\xi_i^2 m^2/|t_i|\ll 1$ and we can keep only the first term of the
expansion~(\ref{legendre}). The tensor product is given by
\begin{eqnarray}
\label{R0p}
&&T^{J_1}(p_1,\Delta_1)\otimes F^{J_1,\; J_2,\; 0^{+}}(\Delta_1,\Delta_2)\otimes T^{J_2}(p_2,\Delta_2)\sim\\
&&\sim s_1^{J_1} s_2^{J_2} \sum_{k=0}^{J_1} \frac{f_k 2^k}{M_{\bot}^{2k}}\; .\nonumber
\end{eqnarray}

\item $J^{P}=0^{-}$
\begin{eqnarray}
\label{J0m}
&&F^{\mu_1\dots\mu_{J_1},\;\nu_1\dots\nu_{J_2}}(\Delta_1,\Delta_2,p_X)= Q_n^{\mu_1\nu_1} \sum_{k=0}^{J_1-1}
f_k\cdot\\
&&\cdot \left( p_X^{\mu_2}\cdot\dots\cdot p_X^{\mu_{J_1-k}}\cdot 
g^{\mu_{J_1-k+1}\nu_{J_2-k+1}}\cdot\dots\cdot g^{\mu_{J_1}\nu_{J_2}}\cdot 
p_X^{\nu_{J_2-k}}\cdot\dots\cdot p_X^{\nu_2}\right)\;,\nonumber
\end{eqnarray}
\begin{eqnarray}
\label{R0m}
&&T^{J_1}(p_1,\Delta_1)\otimes F^{J_1,\; J_2,\; 0^{-}}(\Delta_1,\Delta_2)\otimes T^{J_2}(p_2,\Delta_2)\sim\\
&&\sim 4 Q_n^{\mu_1\nu_1} p_1^{\mu_1} p_2^{\nu_1} s_1^{J_1-1} s_2^{J_2-1} \sum_{k=0}^{J_1-1} \frac{f_k 2^k}{M_{\bot}^{2k}}\; \nonumber\\
&&\simeq \left[ {\bf\Delta}_1\times{\bf\Delta}_2\right]\cdot s_1^{J_1} s_2^{J_2} \sum_{k=0}^{J_1-1} \frac{f_k 2^{k+1}}{M_{\bot}^{2k+2}}\; \nonumber
\end{eqnarray}

\item $J^{P}=1^{-}$
\begin{eqnarray}
\label{J1m}
&&F^{\mu_1\dots\mu_{J_1},\;\nu_1\dots\nu_{J_2},\;\alpha}(\Delta_1,\Delta_2,p_X)= 
g^{\alpha\mu_1} \sum_{k=0}^{J_1-1}
f_k\cdot\\
&&\cdot \left( p_X^{\mu_2}\cdot\dots\cdot p_X^{\mu_{J_1-k}}\cdot 
g^{\mu_{J_1-k+1}\nu_{J_2-k+1}}\cdot\dots\cdot g^{\mu_{J_1}\nu_{J_2}}\cdot 
p_X^{\nu_{J_2-k}}\cdot\dots\cdot p_X^{\nu_1}\right)+\nonumber\\
&&+g^{\alpha\nu_1} \sum_{k=0}^{J_1}
f_{J_1+k}\cdot\nonumber\\
&&\cdot \left( p_X^{\mu_1}\cdot\dots\cdot p_X^{\mu_{J_1-k}}\cdot 
g^{\mu_{J_1-k+1}\nu_{J_2-k+1}}\cdot\dots\cdot g^{\mu_{J_1}\nu_{J_2}}\cdot 
p_X^{\nu_{J_2-k}}\cdot\dots\cdot p_X^{\nu_2}\right)+\nonumber\\
&&+\Delta_n^{\alpha} \sum_{k=0}^{J_1}
f_{2J_1+k+1}\cdot\nonumber\\
&&\cdot \left( p_X^{\mu_1}\cdot\dots\cdot p_X^{\mu_{J_1-k}}\cdot 
g^{\mu_{J_1-k+1}\nu_{J_2-k+1}}\cdot\dots\cdot g^{\mu_{J_1}\nu_{J_2}}\cdot 
p_X^{\nu_{J_2-k}}\cdot\dots\cdot p_X^{\nu_1}\right)\; ,\nonumber
\end{eqnarray}
\begin{eqnarray}
\label{R1m}
&&T^{J_1}(p_1,\Delta_1)\otimes F^{J_1,\; J_2,\; 1^{-}}(\Delta_1,\Delta_2)\otimes T^{J_2}(p_2,\Delta_2)\sim\\
&&\sim s_1^{J_1} s_2^{J_2} \left[ \frac{2p_1^{\alpha}}{s_1}
\sum_{k=0}^{J_1-1} \frac{f_k 2^k}{M_{\bot}^{2k}}\right. +\; \nonumber\\
&&+ \frac{2p_2^{\alpha}}{s_2}
\sum_{k=0}^{J_1} \frac{f_{J_1+k} 2^k}{M_{\bot}^{2k}}+\; \nonumber\\
&&+ \left. \Delta_n^{\alpha}\sum_{k=0}^{J_1} \frac{f_{2J_1+1+k} 2^k}{M_{\bot}^{2k}}\right]\; .\nonumber 
\end{eqnarray}
It is easy to show from the general form of tensor decompositions, that after reggeization procedure one obtains the structure of the amplitude, which is similar to the 
special case $J_1=J_2=1$. In this case it is convenient to use the following bose-symmetric form
of the tensor
\begin{eqnarray}
\label{J1m1}
&&F^{\mu\nu,\;\alpha}(\Delta_1,\Delta_2,p_X)= 
f_0g^{\alpha\mu}\Delta_1^{\nu}+{\bar f}_0g^{\alpha\nu}\Delta_2^{\mu}+\\
&&+\left(f_1\Delta_1^{\alpha}+{\bar f}_1\Delta_2^{\alpha}\right)\Delta_2^{\mu}\Delta_1^{\nu}+\nonumber\\
&&+\left(f_2\Delta_1^{\alpha}+{\bar f}_2\Delta_2^{\alpha}\right)g^{\mu\nu}\;,\nonumber
\end{eqnarray}
and the tensor product
\begin{eqnarray}
\label{R1m1}
&&T^{J_1}(p_1,\Delta_1)\otimes F^{J_1,\; J_2,\; 1^{-}}(\Delta_1,\Delta_2)\otimes T^{J_2}(p_2,\Delta_2)\sim\\
&&\sim s_1^{J_1} s_2^{J_2} \left[ \frac{2p_1^{\alpha}}{s_1}f_0+
\frac{2p_2^{\alpha}}{s_2}{\bar f}_0 +  
\left(f_1+\frac{f_2}{M_{\bot}^2}\right)\Delta_1^{\alpha}+\right.\; \nonumber\\
&&+
\left. \left({\bar f}_1+\frac{{\bar f}_2}{M_{\bot}^2}\right)\Delta_2^{\alpha}
\right]\; ,\nonumber 
\end{eqnarray}
where ${\bar f}(t_1,t_2)=f(t_2,t_1)$.

\item $J^{P}=1^{+}$
\begin{eqnarray}
\label{J1p}
&&F^{\mu_1\dots\mu_{J_1},\;\nu_1\dots\nu_{J_2},\;\alpha}(\Delta_1,\Delta_2,p_X)= 
\Lambda_X^{\mu_1\nu_1\alpha} \sum_{k=0}^{J_1-1}
f_k\cdot\\
&&\cdot \left( p_X^{\mu_2}\cdot\dots\cdot p_X^{\mu_{J_1-k}}\cdot 
g^{\mu_{J_1-k+1}\nu_{J_2-k+1}}\cdot\dots\cdot g^{\mu_{J_1}\nu_{J_2}}\cdot 
p_X^{\nu_{J_2-k}}\cdot\dots\cdot p_X^{\nu_2}\right)+\nonumber\\
&&+\Lambda_n^{\mu_1\nu_1\alpha} \sum_{k=0}^{J_1-1}
f_{J_1+k}\cdot\nonumber\\
&&\cdot \left( p_X^{\mu_2}\cdot\dots\cdot p_X^{\mu_{J_1-k}}\cdot 
g^{\mu_{J_1-k+1}\nu_{J_2-k+1}}\cdot\dots\cdot g^{\mu_{J_1}\nu_{J_2}}\cdot 
p_X^{\nu_{J_2-k}}\cdot\dots\cdot p_X^{\nu_2}\right)+\nonumber\\
&&+Q_n^{\nu_1\alpha}f_{2J_1}\left(
g^{\mu_1\nu_2}\cdot\dots\cdot g^{\mu_{J_1}\nu_{J_1+1}}\cdot 
p_X^{\nu_{J_1+2}}\cdot\dots\cdot p_X^{\nu_{J_2}}\right)+\nonumber\\ 
&&+Q_n^{\mu_1\alpha} \sum_{k=0}^{J_1-1}
f_{2J_1+k+1}\cdot\nonumber\\
&&\cdot \left( p_X^{\mu_2}\cdot\dots\cdot p_X^{\mu_{J_1-k}}\cdot 
g^{\mu_{J_1-k+1}\nu_{J_2-k+1}}\cdot\dots\cdot g^{\mu_{J_1}\nu_{J_2}}\cdot 
p_X^{\nu_{J_2-k}}\cdot\dots\cdot p_X^{\nu_1}\right)\; ,\nonumber
\end{eqnarray}
\begin{eqnarray}
\label{R1p}
&&T^{J_1}(p_1,\Delta_1)\otimes F^{J_1,\; J_2,\; 1^{+}}(\Delta_1,\Delta_2)\otimes T^{J_2}(p_2,\Delta_2)\sim\\
&&\sim s_1^{J_1} s_2^{J_2} 
\left[ \frac{2p_1^{\mu_1}p_2^{\nu_1}p_X^{\rho}\epsilon^{\rho\alpha\mu_1\nu_1}}{s}
\sum_{k=0}^{J_1-1} \frac{f_k 2^{k+1}}{M_{\bot}^{2k+2}}\right. +\; \nonumber\\
&&+ \frac{2p_1^{\mu_1}p_2^{\nu_1}\Delta_n^{\rho}\epsilon^{\rho\alpha\mu_1\nu_1}}{s}
\sum_{k=0}^{J_1-1} \frac{f_{J_1+k} 2^{k+1}}{M_{\bot}^{2k+2}}+\; \nonumber\\
&&+ \frac{2p_2^{\nu_1}p_X^{\rho}\Delta_n^{\sigma}\epsilon^{\rho\sigma\alpha\nu_1}}{s_2}
\frac{f_{2J_1} 2^{J_1}}{M_{\bot}^{2J_1}}+\; \nonumber\\
&&+\left.\frac{2p_1^{\mu_1}p_X^{\rho}\Delta_n^{\sigma}\epsilon^{\rho\sigma\alpha\mu_1}}{s_1}
\sum_{k=0}^{J_1-1}\frac{f_{2J_1+1+k} 2^{k}}{M_{\bot}^{2k}}\; \right]\; .\nonumber 
\end{eqnarray}
For $J_1=J_2=1$
\begin{eqnarray}
\label{R1p1}
&&T^{J_1}(p_1,\Delta_1)\otimes F^{J_1,\; J_2,\; 1^{+}}(\Delta_1,\Delta_2)\otimes T^{J_2}(p_2,\Delta_2)\sim\\
&&\sim s_1^{J_1} s_2^{J_2} \left[ \frac{4p_1^{\sigma}p_2^{\lambda}
\left(f_0\Delta_1^{\rho}-{\bar f}_0\Delta_2^{\rho}
\right)}{s M_{\bot}^2}+ 
2\Delta_1^{\rho}\Delta_2^{\sigma}\left(
\frac{f_1 p_1^{\lambda}}{s_1}-\frac{{\bar f}_1 p_2^{\lambda}}{s_2}
\right)
\right]\epsilon^{\rho\sigma\lambda\alpha}=\; \nonumber\\
&&= 2 s_1^{J_1} s_2^{J_2} \left[ p_1^{\lambda}P_2^{\rho}\Delta_1^{\sigma}+
p_2^{\lambda}P_1^{\rho}\Delta_2^{\sigma}
\right]\epsilon^{\rho\sigma\lambda\alpha}
\;,\nonumber\\
&& P_1=-\frac{{\bar f}_1}{s_2}\Delta_1+\frac{{\bar f}_0}{s M_{\bot}^2}2p_1\;,\\
&& P_2=-\frac{f_1}{s_1}\Delta_2+\frac{f_0}{s M_{\bot}^2}2p_2\;.\nonumber
\end{eqnarray}

\item $J^{P}=2^{+}$. For simplicity we consider only the case $J_1=J_2=1$.
\begin{eqnarray}
\label{J2p}
&&F^{\mu\nu,\;\alpha_1\alpha_2}(\Delta_1,\Delta_2,p_X)= 
f_0g^{\alpha_1\mu}g^{\alpha_2\nu}+\\
&&+f_1g^{\alpha_1\mu}\Delta_1^{\alpha_2}\Delta_1^{\nu}+{\bar f}_1g^{\alpha_1\nu}\Delta_2^{\alpha_2}\Delta_2^{\mu}+
\left( f_2\Delta_1^{\alpha_1}\Delta_1^{\alpha_2}+
{\bar f}_2\Delta_2^{\alpha_1}\Delta_2^{\alpha_2}\right)g^{\mu\nu}+\nonumber\\
&&+\left( f_3\Delta_1^{\alpha_1}\Delta_1^{\alpha_2}+
{\bar f}_3\Delta_2^{\alpha_1}\Delta_2^{\alpha_2}\right)\Delta_2^{\mu}\Delta_1^{\nu}\; ,\nonumber
\end{eqnarray}
\begin{eqnarray}
\label{R2p}
&&T^{J_1}(p_1,\Delta_1)\otimes F^{J_1,\; J_2,\; 2^{+}}(\Delta_1,\Delta_2)\otimes T^{J_2}(p_2,\Delta_2)\sim\\
&&\sim s_1^{J_1} s_2^{J_2} \Bigl[ f_0\frac{4 p_1^{\alpha_1}p_2^{\alpha_2}}{s M_{\bot}^2}+
f_1\frac{2 p_1^{\alpha_1}\Delta_1^{\alpha_2}}{s_1}+
{\bar f}_1\frac{2 p_2^{\alpha_1}\Delta_2^{\alpha_2}}{s_2}+ \Bigr.\nonumber\\
&&+ \Bigl. \frac{2 \left( f_2\Delta_1^{\alpha_1}\Delta_1^{\alpha_2}+
{\bar f}_2\Delta_2^{\alpha_1}\Delta_2^{\alpha_2}\right)}{M_{\bot}^2}
+ \left( f_3\Delta_1^{\alpha_1}\Delta_1^{\alpha_2}+
{\bar f}_3\Delta_2^{\alpha_1}\Delta_2^{\alpha_2}\right)\Bigr]\; .\nonumber 
\end{eqnarray}

\end{itemize}

Everywhere in the above expressions $f_k=f^{J^P}_k(t_1,t_2,M_X^2)$.

\section{Azimuthal angle dependence.}

In this section we will obtain the general structure of
the azimuthal angle dependence for different $J^P$ states in the
tensor current picture. If we assume that the dominant
contribution is given by Regge poles $\alpha_1(t_1)$ and $\alpha_2(t_2)$ then the amplitude 
of the process can be written in the following form
\begin{equation}
\label{TXgeneral}
T_{p p\to p X p}\sim\eta(\alpha_1(t_1))\eta(\alpha_2(t_2)) 
\left[ T^{J_1}\otimes F^{J_1,\; J_2,\; J^P}\otimes T^{J_2}\right]_{{J_1\to \alpha_1}\atop{J_2\to\alpha_2}}\;,
\end{equation}
where $J_i\to\alpha_i$ means the usual procedure of the analytical continuation
to the complex $J$-plane and taking residues 
at Regge poles, $\eta(\alpha_i)$ are signature factors.

In the most important case, when the main contribution
is given by one Pomeron trajectory, $\alpha_1=\alpha_2=\alpha_{\mathbb P}(0)$ and the "bare" amplitude 
squared is proportional to following expressions:

\begin{itemize}
\item $J^P=0^{+}$
\begin{equation}
\label{fi0p}
|T_{p p\to p X p}|^2\sim (M_{\bot}^2)^{2(\alpha_{\mathbb P}(0)-1)}(f_0M_{\bot}^2+2f_1)^2
\end{equation}

\item $J^P=0^{-}$
\begin{equation}
\label{fi0m}
|T_{p p\to p X p}|^2\sim (M_{\bot}^2)^{2(\alpha_{\mathbb P}(0)-1)}f_0 t_1 t_2 \sin^2\phi_0
\end{equation}

\item $J^P=1^{-}$. In this case we assume the existence of a C-odd vacuum trajectory, "Odderon", $\alpha_{\mathbb O}(t)$.
\begin{equation}
\label{fi1m}
|T_{p p\to p X p}|^2\sim (M_{\bot}^2)^{\alpha_{\mathbb P}(0)+\alpha_{\mathbb O}(0)-2}({\cal F}_0 M_{\bot}^4+{\cal F}_1 M_{\bot}^2+{\cal F}_2)\;,
\end{equation}
\begin{eqnarray}
\label{funJ1m}
&& {\cal F}_0=\frac{{f_0^S}^2}{M_X^2}+
f_0^Af_1^A+\frac{(t_1-t_2)f_1^Af_0^S}{M_X^2}+
\frac{{f_1^A}^2\lambda}{4 M_X^2}\nonumber\;,\\
&& {\cal F}_1={f_0^A}^2-{f_0^S}^2-
f_0^Af_2^A+\frac{(t_1-t_2)f_2^Af_0^S}{M_X^2}+
\frac{f_1^Af_2^A\lambda}{2 M_X^2}\nonumber\;,\\
&& {\cal F}_2=\frac{{f_2^A}^2\lambda}{4M_X^2}\;,\\
&&\lambda=\lambda(M_X^2,t_1,t_2)=M_X^4+t_1^2+t_2^2-2M_X^2t_1-2M_X^2t_2-2t_1t_2\;,\nonumber\\
&&f_k^S=f_k+{\bar f}_k\;, f_k^A=f_k-{\bar f}_k\;.\nonumber
\end{eqnarray}

\item $J^P=1^{+}$
\begin{equation}
\label{fi1p}
|T_{p p\to p X p}|^2\sim (M_{\bot}^2)^{2(\alpha_{\mathbb P}(0)-1)}({\cal F}_0 M_{\bot}^4+{\cal F}_1 t_1 t_2 \sin^2\phi_0+{\cal F}_2)\;,
\end{equation}
\begin{eqnarray}
\label{funJ1p}
&& {\cal F}_0=(f_1{\bf\Delta}_1-{\bar f}_1{\bf\Delta}_2)^2\;,\\
&& {\cal F}_1=\frac{2(s_2f_1-s_1{\bar f}_1)^2}{s}+4 M_{\bot}^2 f_1{\bar f}_1-
\frac{4(f_0+{\bar f}_0)^2}{M_X^2}\;,\\
&& {\cal F}_2=4(f_0{\bf\Delta}_1-{\bar f}_0{\bf\Delta}_2)^2\;,
\end{eqnarray}


\item $J^P=2^{+}$
\begin{equation}
\label{fi2p}
|T_{p p\to p X p}|^2\sim (M_{\bot}^2)^{2(\alpha_{\mathbb P}(0)-1)}({\cal F}_0 M_{\bot}^4+{\cal F}_1 M_{\bot}^2+{\cal F}_2)\;,
\end{equation}
\begin{eqnarray}
\label{funJ2p}
&& {\cal F}_0=\frac{{f_1^S}^2-12f_0f_3^S}{24}+\frac{f_1^S(-4f_0+f_1^A(t_1-t_2)+2\lambda f_3^S)}{12M_X^2}+\\
&&+ \frac{1}{24M_X^4}\left[
16f_0^2+4f_0\left( 4 f_1^A(t_1-t_2)+f_3^S(3(t_1-t_2)^2-\lambda)\right)+\right.\nonumber\\
&&+\left. 4f_3^Sf_1^A(t_1-t_2)\lambda+{f_3^S}^2\lambda+{f_1^A}^2\left( (t_1-t_2)^2+3\lambda\right)
\right]\nonumber\;,\\
&& {\cal F}_1=-\frac{1}{3}f_0\left( f_1^S+3f_2^S \right)+\\
&&+\frac{f_2^S}{6M_X^4}\left[ \lambda^2 f_3^S+2\lambda\left(-f_0+f_1^A(t_1-t_2)\right)+
6f_0(t_1-t_2)^2\right]+\nonumber\\
&&+ \frac{1}{6 M_X^2}\left[
\lambda\left( 2f_1^Sf_2^S+2f_0f_3^S+3({f_1^S}^2-{f_1^A}^2)/4\right)-
4f_0^2-2f_0f_1^A(t_1-t_2)
\right]\nonumber\;,\\
&& {\cal F}_2=\frac{(2f_0+f_2^S\lambda /M_X^2)^2}{6}\;.
\end{eqnarray}

\end{itemize}

Similar formulae for the differential cross-sections were
obtained by other authors. In Ref.~\cite{Close} results
were obtained from the assumption that the Pomeron acts as a $1^+$ 
conserved or nonconserved current. It was shown 
in Ref.~\cite{Khoze} and with more detailed investigations in Ref.~\cite{Khoze2} that the same
result follows from the simple Regge behaviour of helicity
amplitudes. Experimental data are in good agreement
with these predictions.

A good description of those data in the framework of our approach (with account of the absorbtion) is obtained (Figs. 3,4). 

In addition to the present description there are some models of "glueball" based on the "instanton" dynamics~\cite{instanton}. 

\section{WA102 and predictions for LHC.}

 It is important to stress the fact, that at WA102 energies
absorbtive effects are
not so significant, and azimuthal angle dependence
looks like the "bare" one. We can use this fact to simplify the fitting procedure, that has been already done by WA102 collaboration. Only at large values of ${\rm dP}_{\bot}=|{\bf\Delta}_1-{\bf\Delta}_2|$ the process of "soft" rescattering can change the picture violently (see Fig.~3d). 

 In Figs.~3,4 we show the data from WA102~\cite{WA102} and our curves for "bare" and unitarized amplitudes. One can see that all the features of the average $\phi_0$-dependence are consistent with the data. It makes possible to predict azimuthal angle behaviour at higher energies and use these predictions as a spin-parity analyser. 
 
 The main property is that unitarization adds up to the distortion of the "bare" cross-section towards small angles and to the reduction of its value. The difference is more significant at LHC than at WA102 energies, and we should take it into account necessarily. "Soft" survival probability is $0.25\to 0.3$ for WA102 and $0.05\to 0.1$ for LHC. It depends on mass $M_X$ and kinematical cuts. 
 
 Features concerning each particle are the same as was mentioned in Ref.~\cite{Close}:
 
 \begin{itemize}
\item for $\eta^{\prime}$ mesons the kinematical distortion because of the different reference frame is totally compensated by unitarization at WA102 energies (Fig. 3a), and for LHC the peak in the cross-section is shifted to $~65^o$ (Fig. 5a). 
\item for ${\rm f}_1(1285)$ we have almost "flat" distribution at large values of $|t_1-t_2|$ (Fig. 3d), since in the simplest case its cross-section is proportional to ${\rm dP}_{\bot}^2=({\bf\Delta}_1-{\bf\Delta}_2)^2$. For LHC we obtain more stronger suppression for large angles (Fig. 5d).
\item the difference between $0^{++}$ $q{\bar q}$ and non-$q{\bar q}$ states at WA102 (Figs. 4a,b correspondingly) changes due to unitarization effects. For $q{\bar q}$ azimuthal angle dependence becomes almost "flat" (Fig. 6b), and for non-$q{\bar q}$ mesons we see the shrinkage of the peak at $\phi_0=0^o$. The same is valid for $2^{++}$ mesons.
\end{itemize}
 
 For low mass particles, which can be produced in the EDDE, total cross-sections are estimated to be of the order $1\div 30\;\mu b$ at LHC. Cross-sections for "glueball" candidates ${\rm f}_0(1500)$ and ${\rm f}_2(1950)$ are about $30\;\mu b$ (depend on the LHC kinematics and may be larger) and effective slope is $~10$. Typical values of $\xi$ are of the order$M_X/\sqrt{s}\sim 10^{-4}$. 
 
 It is possible to apply the method to large mass particles. In Ref.~\cite{Cho} it was argued that a heavy glueball ("knot") $1^{--}$ with mass near $50$~GeV may exist. One can see from~(\ref{fi0p})-(\ref{fi2p}) that in this case $\phi_0$ dependence is simply defined by unitarization only, since\linebreak $M^2_{\bot}\simeq M^2_X=const$. In this case we have a good tool to check models for "soft" rescattering. Examples are depicted in Fig.7.
 
\section*{Conclusions}
 
Detailed investigations of the azimuthal angle dependence in the EDDE can help to solve several important problems:
\begin{itemize}
\item to check different models for "soft" processes and to study the real pattern of the interaction. 
\item to understand the difference in the dynamics of production of $q{\bar q}$ and non-$q{\bar q}$ states and their possible filtering. 
\item to determine the quantum numbers of new produced particles.
\end{itemize}

Experimental possibilities of angular distributions measurements should be studied in further Monte Carlo simulations. 

\section*{Aknowledgements}

The work is supported by grants SNRS-PICS-2910 and RFBR-04-02-17299.
 



\newpage
\section*{Figure captions}

\begin{list}{Fig.}{}

\item 1: The process $p+p\to p+X+p$. Absorbtion in the initial and final pp-channels is not shown.
\item 2: The full unitarization of the process $p+p\to p+X+p$.
\item 3: Experimental data from WA102. Dashed curve represents "bare" cross-section and solid one is the unitarized result. a) $\eta^{\prime},\; 0^{-+}$; b) $f_1(1285),\; 1^{++}$, all $t_i$; c) $f_1(1285)$, $|t_1-t_2|<0.2\; {\rm GeV}^2$; d) $f_1(1285)$, $|t_1-t_2|>0.4\; {\rm GeV}^2$;

\item 4: Experimental data from WA102, averaged over all measured $t_i$ values. Dashed curve represents "bare" cross-section and solid one is the unitarized result.\linebreak a)$f_0(980),\; 0^{++}$; b)$f_0(1500),\; 0^{++}$; c)$f_2(1270),\; 2^{++}$; d) $f_2(1950),\; 2^{++}$;

\item 5: Results for LHC energies. a) $\eta^{\prime},\; 0^{-+}$; b) $f_1(1285),\; 1^{++}$, all $t_i$; c) $f_1(1285)$, $|t_1-t_2|<0.1\; {\rm GeV}^2$; d) $f_1(1285)$, $|t_1-t_2|>0.2\; {\rm GeV}^2$;

\item 6: Results for LHC energies. a)$f_0(980),\; 0^{++}$; b)$f_0(1500),\; 0^{++}$; c)$f_2(1270),\; 2^{++}$; d) $f_2(1950),\; 2^{++}$;

\item 7: Examples of azimuthal angle dependence for large mass states. a) $1^{--}$, $M_X=50\;{\rm GeV}$; b) $0^{-+}$, $M_X=50\;{\rm GeV}$;

\end{list}


\newpage

\begin{figure}[hb]
\label{bare}
\vskip 4cm
\hskip  1cm \vbox to 14cm {\hbox to 16cm{\epsfxsize=16cm\epsfysize=14cm\epsffile{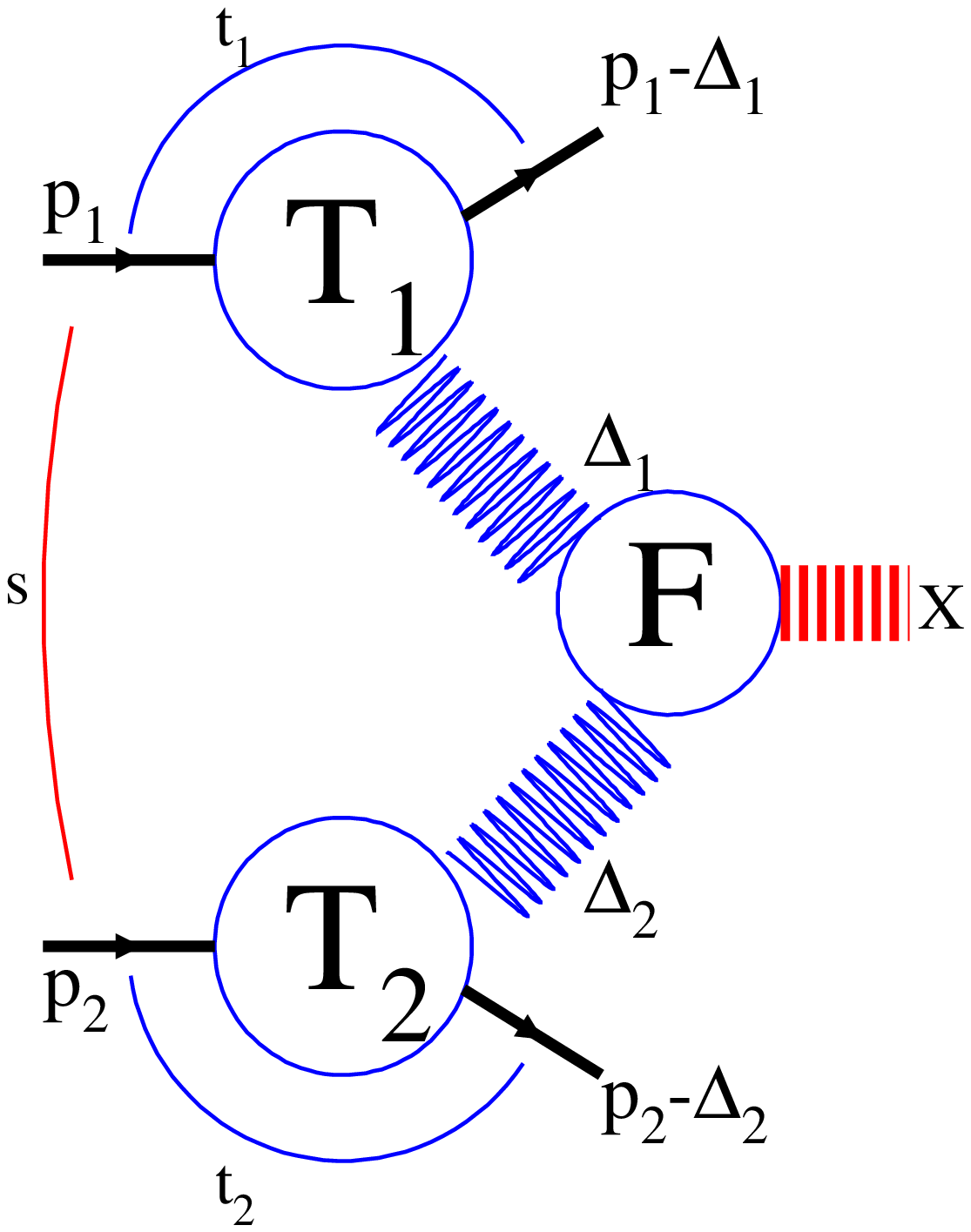}}}
\hskip 1cm
\caption{}
\end{figure}

\newpage

\begin{figure}[hb]
\label{unitar}
\vskip 1.5cm
\vbox to 15cm {\hbox to 15cm{\epsfxsize=15cm\epsfysize=15cm\epsffile{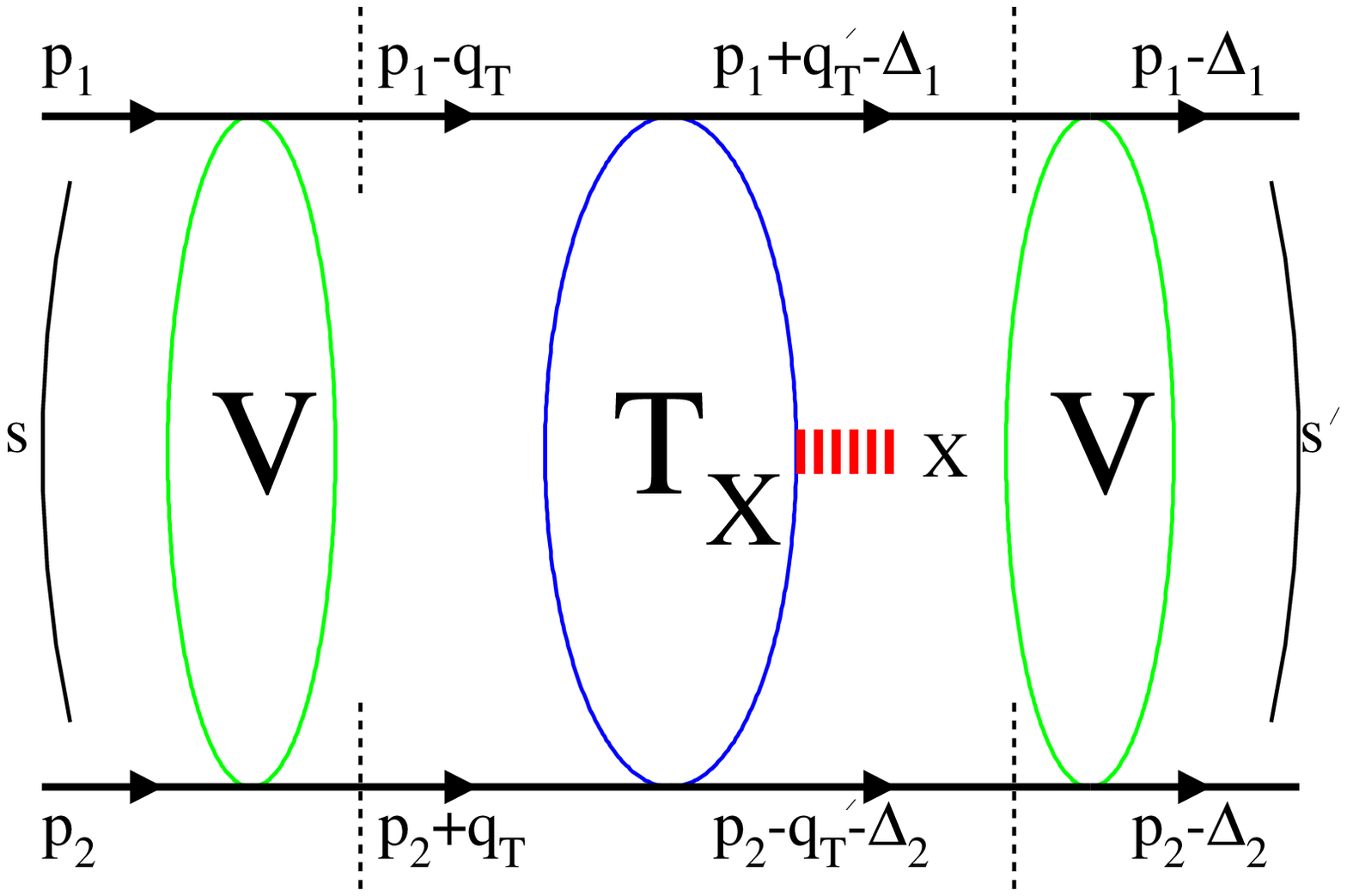}}}
\hskip 1cm
\caption{}
\end{figure}

\newpage

\begin{figure}[h]
\label{WA102A}
\mbox{a)}\hskip -0.5cm
\vbox to 7cm {\hbox to 7cm{\epsfxsize=7cm\epsfysize=7cm\epsffile{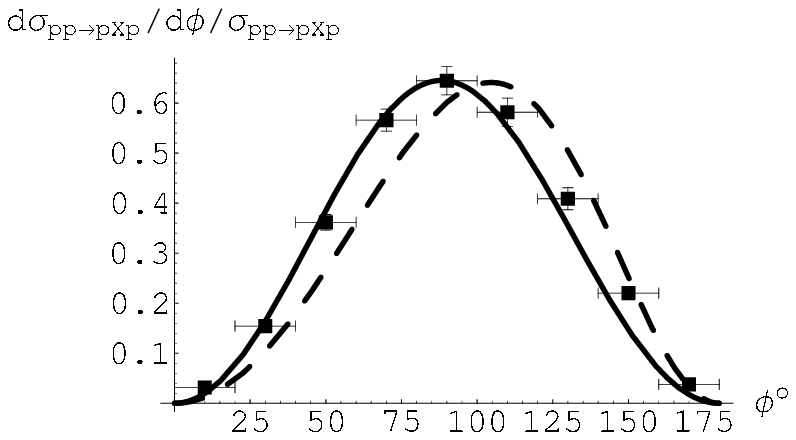}}}
\mbox{b)}
\vskip -7cm
\hskip 7cm
\vbox to 7cm {\hbox to 7cm{\epsfxsize=7cm\epsfysize=7cm\epsffile{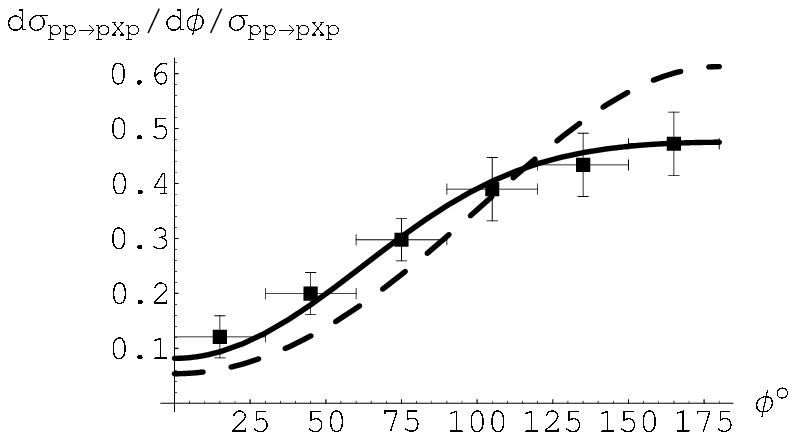}}}
\\
\mbox{c)}\hskip -0.5cm
\vbox to 7cm {\hbox to 7cm{\epsfxsize=7cm\epsfysize=7cm\epsffile{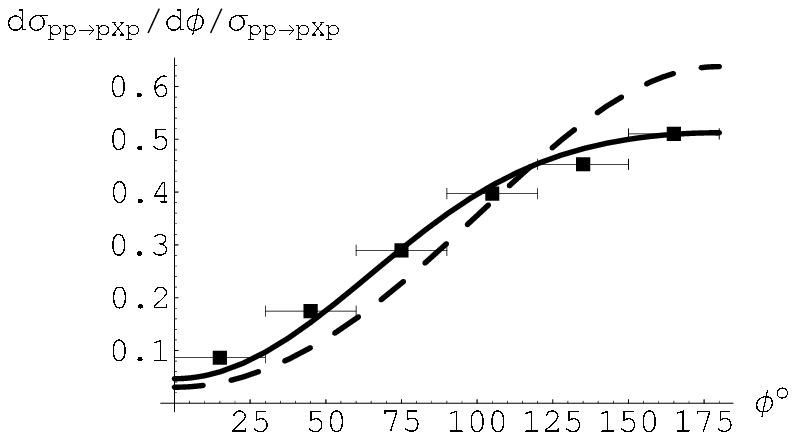}}}
\vskip -7cm
\hskip 7cm
\mbox{d)}
\vbox to 7cm {\hbox to 7cm{\epsfxsize=7cm\epsfysize=7cm\epsffile{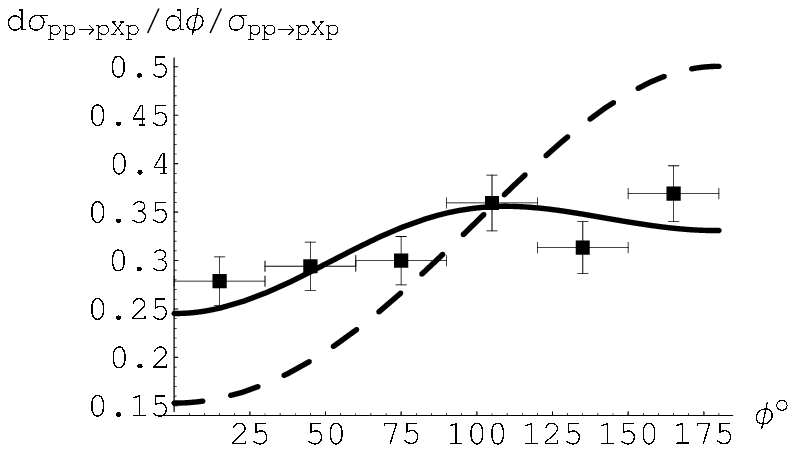}}}
\caption{}
\end{figure}

\newpage

\begin{figure}[h]
\label{WA102B}
\mbox{a)}\hskip -0.5cm
\vbox to 7cm {\hbox to 7cm{\epsfxsize=7cm\epsfysize=7cm\epsffile{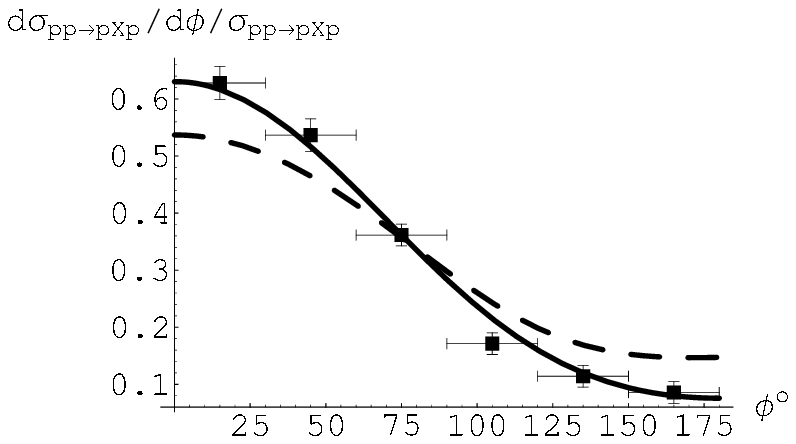}}}
\mbox{b)}
\vskip -7cm
\hskip 7cm
\vbox to 7cm {\hbox to 7cm{\epsfxsize=7cm\epsfysize=7cm\epsffile{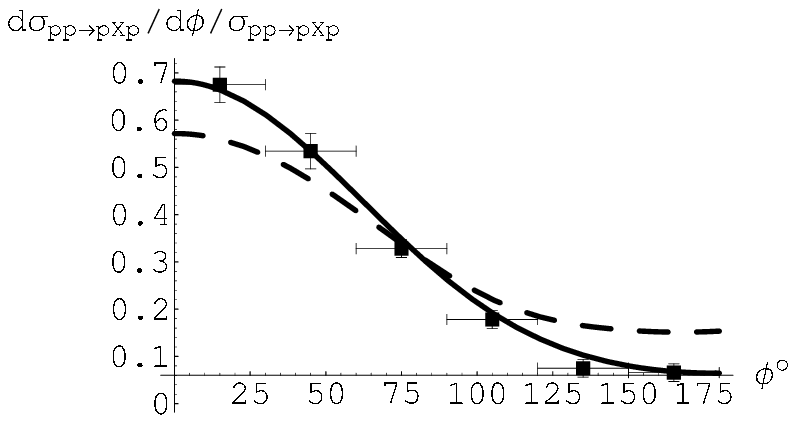}}}
\\
\mbox{c)}\hskip -0.5cm
\vbox to 7cm {\hbox to 7cm{\epsfxsize=7cm\epsfysize=7cm\epsffile{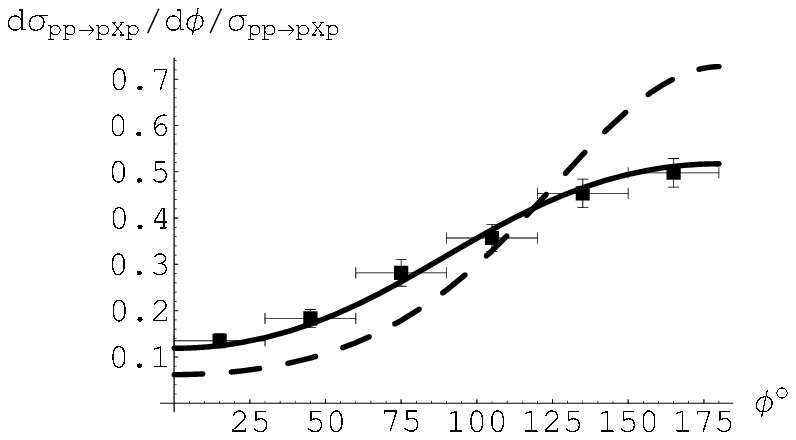}}}
\vskip -7cm
\hskip 7cm
\mbox{d)}
\vbox to 7cm {\hbox to 7cm{\epsfxsize=7cm\epsfysize=7cm\epsffile{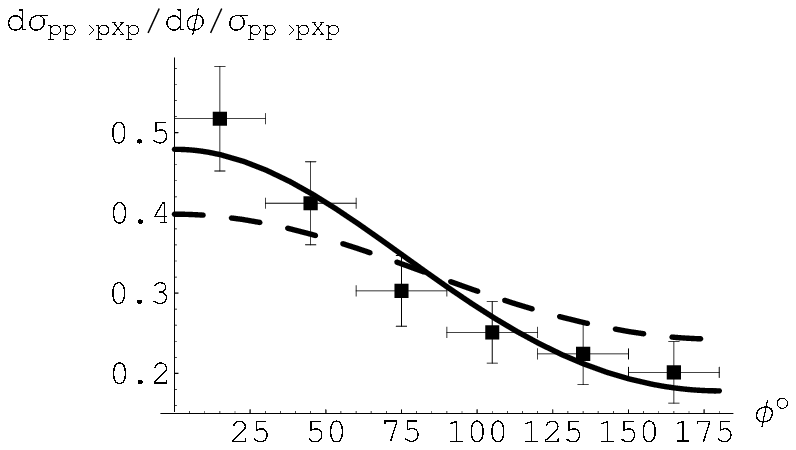}}}
\caption{}
\end{figure}

\newpage

\begin{figure}[h]
\label{LHCA}
\mbox{a)}\hskip -0.5cm
\vbox to 7cm {\hbox to 7cm{\epsfxsize=7cm\epsfysize=7cm\epsffile{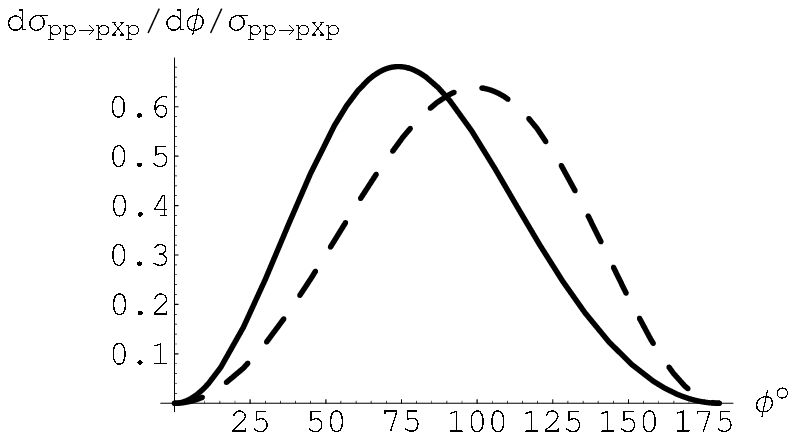}}}
\mbox{b)}
\vskip -7cm
\hskip 7cm
\vbox to 7cm {\hbox to 7cm{\epsfxsize=7cm\epsfysize=7cm\epsffile{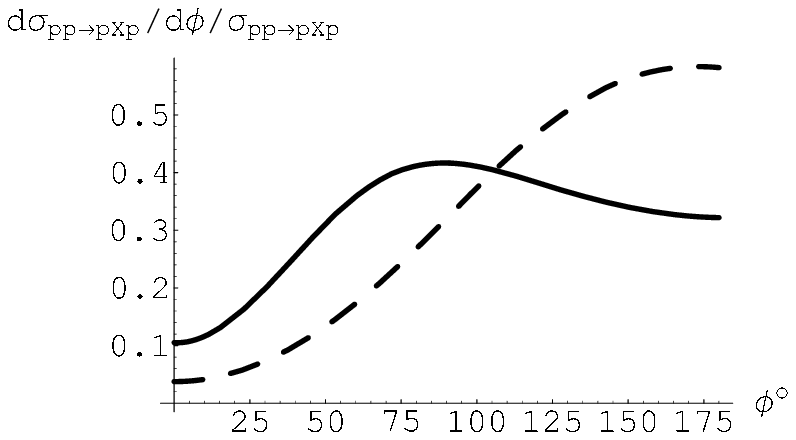}}}
\\
\mbox{c)}\hskip -0.5cm
\vbox to 7cm {\hbox to 7cm{\epsfxsize=7cm\epsfysize=7cm\epsffile{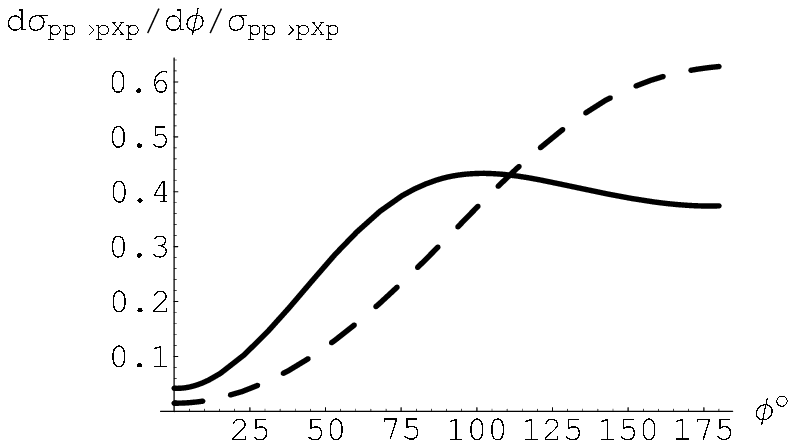}}}
\vskip -7cm
\hskip 7cm
\mbox{d)}
\vbox to 7cm {\hbox to 7cm{\epsfxsize=7cm\epsfysize=7cm\epsffile{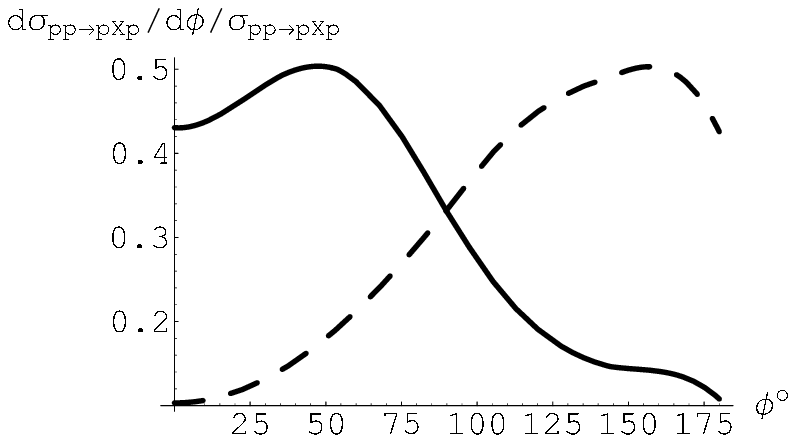}}}
\caption{}
\end{figure}

\newpage

\begin{figure}[h]
\label{LHCB}
\mbox{a)}\hskip -0.5cm
\vbox to 7cm {\hbox to 7cm{\epsfxsize=7cm\epsfysize=7cm\epsffile{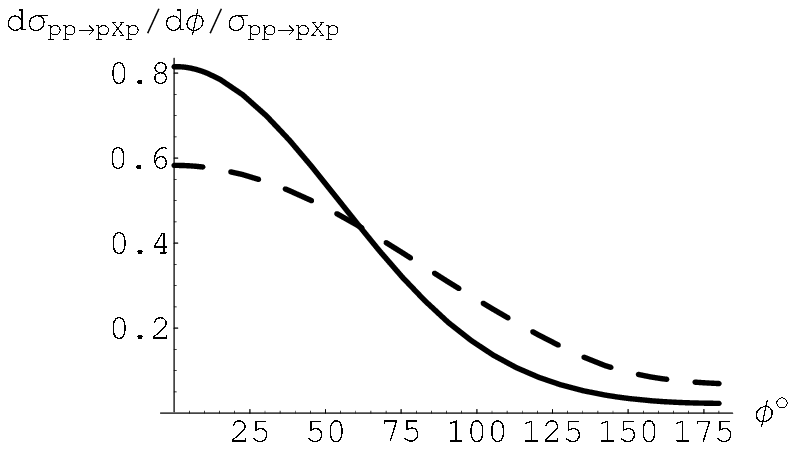}}}
\mbox{b)}
\vskip -7cm
\hskip 7cm
\vbox to 7cm {\hbox to 7cm{\epsfxsize=7cm\epsfysize=7cm\epsffile{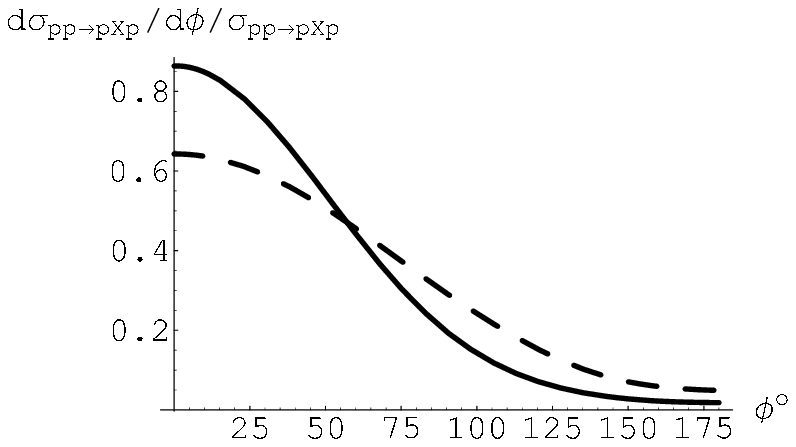}}}
\\
\mbox{c)}\hskip -0.5cm
\vbox to 7cm {\hbox to 7cm{\epsfxsize=7cm\epsfysize=7cm\epsffile{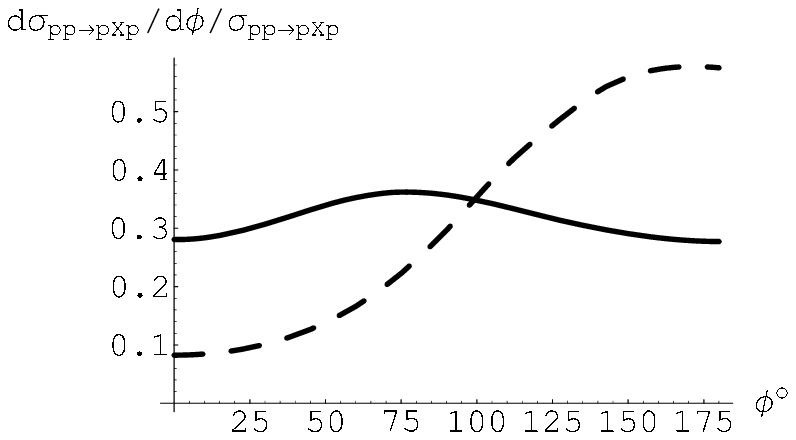}}}
\vskip -7cm
\hskip 7cm
\mbox{d)}
\vbox to 7cm {\hbox to 7cm{\epsfxsize=7cm\epsfysize=7cm\epsffile{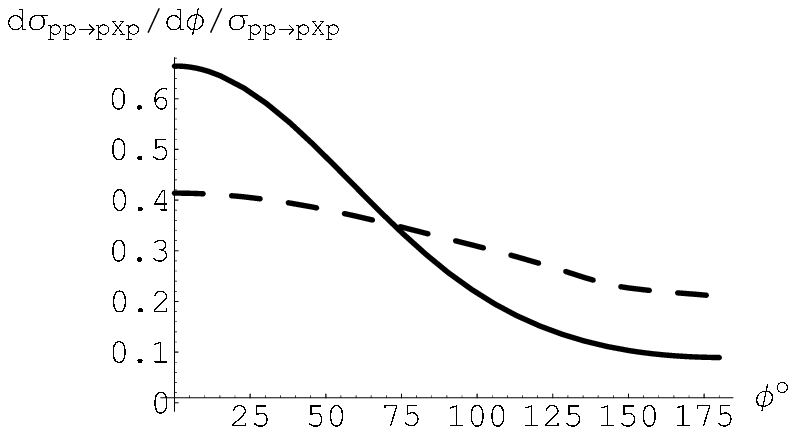}}}
\caption{}
\end{figure}

\newpage

\begin{figure}[h]
\label{LHCMlarge}
\mbox{a)}\hskip -0.5cm
\vbox to 7cm {\hbox to 7cm{\epsfxsize=7cm\epsfysize=7cm\epsffile{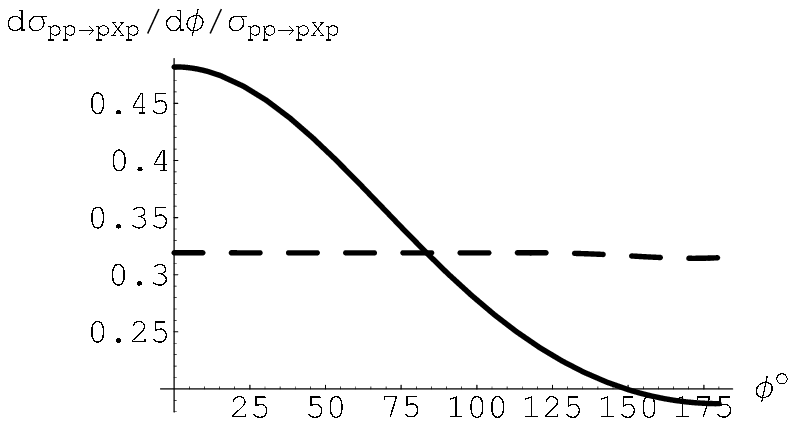}}}
\mbox{b)}
\vskip -7cm
\hskip 7cm
\vbox to 7cm {\hbox to 7cm{\epsfxsize=7cm\epsfysize=7cm\epsffile{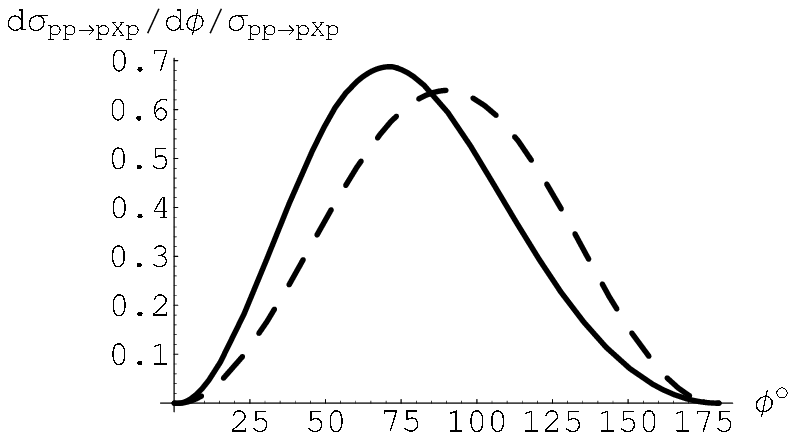}}}
\caption{}
\end{figure}


\begin{thebibliography}{9}

  \bibitem {Logunov} T.W. Kibble, {\it Proc. Roy. Soc. 244 (1958) 355.}\\
  A.A. Logunov and A.N. Tavkhelidze, {\it Nucl. Phys. 8 (1958) 374.}
  \bibitem {2} K.A. Ter-Martirosyan, {\it Nucl. Phys. 68 (1964) 591};\\
  K.G. Boreskov, {\it Yad. Fiz. V.8 S.4 (1968) 796}.
  \bibitem {3} V.A. Petrov and R.A. Ryutin, {\it JHEP 0408 (2004) 013};\\ 
  V.A. Petrov and R.A. Ryutin, {\it Eur. Phys. Journ. C 36 (2004) 509}. 
  \bibitem {4} F.E. Close, A. Kirk, {\it Eur. Phys. J. C 21 (2001) 531}, hep-ph/0103173.
  \bibitem{5} S.S. Gershtein and  A.A. Logunov, {\it Sov. J. Nucl. Phys. 39 (1984) 1514}.  
  \bibitem{Prokoshkin} Yu.D. Prokoshkin, IHEP preprint 88-40. Serpukhov, 1988.
  \bibitem {WA102} WA102 Coll., {\it Phys. Lett. B 427 (1998) 398}, hep-ex/9803029;\\{\it Phys. Lett. B 467 (1999) 165}, hep-ex/9909013;\\           			{\it Phys. Lett. B 474 (2000) 423}, hep-ex/0001017;\\
  \bibitem {WA102a} WA102 Coll., {\it Phys. Lett. B 462 (1999) 462}, hep-ex/9907055.
  \bibitem {WA102b} A. Kirk, {\it Phys. Lett. B 489 (2000) 29}, hep-ph/0008053.
  \bibitem {6} F.E. Close, {\it Phys. Lett. B 419 (1998) 387}, hep-ph/9710450;\\ WA102 Coll., A. Kirk {\it et al.}, hep-ph/9810221;\\  
  F.E. Close, A. Kirk, {\it Phys. Lett. B 397 (1997) 333}, hep-ph/9701222;\\ 
  F.E. Close, A. Kirk, G.A. Schuler, {\it Phys. Lett. B 477 (2000) 13}, hep-ph/0001158.
  \bibitem {Collins} A. Berera and J. C. Collins, {\it Nucl. Phys. B 474 (1996) 183.}
  \bibitem {Cudell} J.-R. Cudell and O. F. Hernandez, {\it Nucl. Phys. B 471 (1996) 471.}
  \bibitem {Petrov3P} V.A. Petrov and A. V. Prokudin, {\it Eur. Phys. J. C 23 (2002)135.}
  \bibitem {Vertex3J} Jie-Jie Zhu, Mu-Lin Yan, hep-ph/9903349. 
  \bibitem {Close} F.E. Close, G.A. Schuller, {\it Phys. Lett. B 458(1999) 127}, hep-ph/9902243;\\ 
  F.E. Close, G.A. Schuller, {\it Phys. Lett. B 464 (1999) 279}, hep-ph/9905305. 
  \bibitem {Khoze} A.B. Kaidalov, V.A. Khoze, A.D. Martin and M.G. Ryskin, {\it Eur. Phys. J. C {\bf 31} (2003) 387}, hep-ph/0307064.
  \bibitem{Khoze2} V.A. Khoze, A.D. Martin and M.G. Ryskin, {\it Eur. Phys. J. C {\bf 24} (2002) 581}, hep-ph/0203122.
  \bibitem{instanton} E. Shuryak, I. Zahed, {\it Phys. Rev. D {\bf 68} (2003) 034001}, hep-ph/0302231;\\
N. Kochelev, hep-ph/9902203

  \bibitem{Cho} Y.M. Cho, hep-th/0406004.
  
 
\end{thebibliography}
\end{document}